\documentclass[twocolumn,english]{revtex4-1}
\usepackage[T1]{fontenc}
\usepackage[latin9]{inputenc}
\usepackage{babel}
\usepackage{amsmath}
\usepackage{amssymb}
\usepackage{graphicx}
\usepackage[unicode=true]
 {hyperref}

\makeatletter

\providecommand{\tabularnewline}{\\}

\makeatother

\begin{document}
\title{Connecting afterglow light curves to the GRB central engine}
\author{Muhammed Diyaddin Ilhan and Kai Schwenzer}
\address{Istanbul University, Science Faculty, Department of Astronomy and
Space Sciences, Beyazit, 34119, Istanbul, Turkey}
\begin{abstract}
Gamma ray burst (GRB) afterglow light curves have the potential to
inform us about presently unobserved stages in the aftermath of a
neutron star merger. Using numerical simulations of short GRB afterglows
we obtain an approximate quantitative connection between key aspects
of the emission mechanism and the shapes of the resulting light curves.
 Employing simple, but efficient, parameterizations of the light
curve based on a broken power law in terms of physical parameters,
fitted to a large dataset of synthetic light curves, we apply basic
machine learning techniques to determine the approximate connection
between key input parameters of the forward shock model and the light
curve parameters. Solving then the inverse problem, we find that the
strength of the central engine can be reasonably accurately estimated
even with very limited information. In particular, merely the position
of jet-break in the on-axis, respectively the maximum in the off-axis
light curve determines the kinetic energy at the tens of percent level.
\end{abstract}
\maketitle

\section{Introduction}

Gamma ray bursts (GRBs) \citep{Klebesadel:1973iq,Zhang:2018ond} are
the most luminous signals in the known universe, and can exceed the
energy released by the sun during its entire lifetime. They are observed
as intense bursts of gamma-rays lasting from a few milliseconds to
a few minutes \citep{Gao:2015lga,Zhang:2018ond}. They have a structured
duration distribution with two pronounced peaks \citep{Kouveliotou:1993yx}
and can be broadly divided into short and long GRBs. While there were
earlier hints \citep{Gehrels:2005qk,Berger:2013jza,Fong:2015oha},
the short GRB170817 has been clearly linked to the seminal gravitational
wave detection GW170817 from neutron star inspiral \citep{TheLIGOScientific:2017qsa,LIGOScientific:2017ync},
as well as to a corresponding kilonova signaling r-process nucleosynthesis
\citep{Metzger:2016pju}. The early stages of neutron star mergers
\citep{Baiotti:2016qnr}, can be sensitive to the properties of strongly
interacting dense matter \citep{Alford:2017rxf}, and since the density
in the contracting merger product steadily rises, they reach the highest
possible densities and therefore could probe novel phases of matter
\citep{Alford:2019oge}. Up to now there are no direct signals from
the early post-merger stages of a neutron star merger event. Therefore,
it is tempting to consider the various subsequent electromagnetic
signals driven by such a central engine. While the GRB prompt emission
is complicated and the detailed emission mechanism is not fully identified
at this point \citep{2015AdAst2015E..22P,Zhang:2018ond}, it is accompanied
by an extended afterglow \citep{Costa:1997obd,vanParadijs:1997wr,Zhang:2018ond},
spanning nearly the entire electromagnetic spectrum and lasting depending
on the band from days to years. The afterglow has been clearly linked
to synchrotron emission from electrons driven by a highly relativistic
outflow \citep{1975NYASA.262..164R}. And an external, weakly magnetized
external forward shock, where the outflow smashes into the interstellar
medium (ISM) and particles are accelerated far away from the central
engine, has been shown to reproduce the key aspects of the afterglow
emission \citep{Wang:2015vpa,Zhang:2018ond}. In addition to information
about the location of the source, the geometric structure of the jet
\citep{Leventis:2013gf,Ryan:2019fhz} and the environment \citep{Gehrels:2009qy},
the rich dataset of GRB afterglows should contain a lot of information
about the emission process and the central engine, resulting from
the merger, that powers the GRB and its afterglow. Unfortunately,
this connection is complicated by the various dynamic processes at
different stages of these rich events extending over large spatial
and temporal intervals \citep{2014ARA&A..52...43B}. Therefore, we
rely here on realistic simulations of the forward shock mechanism
\citep{Ryan:2019fhz} that capture the key aspects of this complicated
physics, and try to abstract the characteristic imprint of the key
parameter determined by the central engine, namely the total isotropic
kinetic energy, on the light curve morphology.

A jet consisting of ultra-relativistic particles and fields \citep{Salafia:2022dkz,2015PhR...561....1K}
can be launched by the gigantic magnetic fields owing to the accretion
disk surrounding the compact remnant in the immediate aftermath of
the merger \citep{Rezzolla:2011da}, which is most likely a black
hole. GRB afterglows can be described within the forward shock model
\citep{Sari:1997qe,vanEerten:2010zh} as the resulting relativistic
jet plows through the surrounding ISM, and these are powered by an
equivalent isotropic kinetic energy $E_{K,\mathrm{iso}}$ between
around $10^{49}\,\mathrm{erg}$ and $10^{54}\,\mathrm{erg}$ provided
by the central engine \citep{Berger:2013jza,Li:2016pes,2015PhR...561....1K}.
The detailed geometry of this jet is not well established so far but
can have a sizable impact on the resulting light curves \citep{Ryan:2019fhz}.
In this work we provide a simple connection between generic observational
properties of the afterglow light curve and input parameters of the
forward shock model using the simplest ``top-hat'' jet geometry
\citep{Sari:1997qe,vanEerten:2010zh}, in which the energy per solid
angle is constant. In addition to the kinetic energy, being the key
input parameter determined by the central engine, these also include
environment parameters like the observation angle or the density of
the ISM, as well as parameters describing the microscopic processes,
like the frequency dependence of the afterglow emission, the latter
being mainly produced through the synchrotron mechanism \citep{Sari:1997qe}.
GRB afterglows can be divided into on-axis events \citep{Zhang:2018ond},
that are seen at an angle $\theta_{\mathrm{obs}}$ smaller than the
opening angle of the jet $\mathrm{\theta_{\mathrm{jet}}}$ ($\theta_{\mathrm{obs}}<\mathrm{\theta_{\mathrm{jet}}}$)
and off-axis events that are seen at large angles ($\mathrm{\theta_{\mathrm{jet}}}<\theta_{\mathrm{obs}}$).
We distinguish these two qualitatively different classes and parametrize
their light curves by simple but generic models that approximate them
in terms of a few physical parameters. These are then correlated to
the mentioned key parameters describing the simulation to obtain the
leading power law relations between them, by employing linear regression
as well as training a basic neutral network.

\section{Light curve parametrization and fitting}

In general, afterglow light curves are complicated and can have a
rich structure \citep{Zhang:2018ond}, including an initial steep
decay ending the prompt phase, a shallow decay/plateau at early times
and potential additional flares. However, the late time behavior,
that can for most afterglows be associated to the external shock emission,
is more generic. At late times, the most prominent feature in the
on-axis GRB afterglow light curves is the jet break, where the light
curve changes to a steeper power law decay, because the Lorentz factor
of accelerated charges decreases so that the emitted radiation is
less an less beamed and eventually becomes isotropic. If the event
is only seen off-axis, as the beam wides more and more of it enters
the line of sight of the observer, which leads to a maximum in the
light curve that depends both on the time of jet break and the observer
angle. 

The light curve describes the time dependence of the spectral flux
density $F\!\left(t\right)$. Following previous work \citep{Zhang:2005fa,Wang:2018deh},
we model these features by a generalized resonant form with power
law coefficients $a$ and $b$ and exponents $\alpha$ and $\beta$
that is smoothed out by a parameter $\nu$ 

\begin{equation}
F\!\left(t\right)=\left(\left(at^{-\alpha}\right)^{\nu}+\left(bt^{\beta}\right)^{\nu}\right)^{-\frac{1}{\nu}}
\end{equation}
which naturally results in a broken power law with asymptotic exponents
$\alpha$ and $-\beta$. at early and late time. Instead of the coefficients
$a$ and $b$, our goal is to replace them by the observable light
curve parameters, namely by the flux at a characteristic time. This
requires to distinguish between the on-axis and the off-axis case,
where the corresponding parameters are the time $t_{\mathrm{jb}}$
and the corresponding spectral flux density $F_{\mathrm{jb}}$ at
jet break, respectively the peak spectral flux density $F_{\mathrm{max}}$
and the corresponding time $t_{\mathrm{max}}$.

In the off-axis case the maximum of the light curve is determined
by the condition $dF\!\left(t_{\mathrm{max}}\right)/dt=0$ which yields
the time $t_{\mathrm{max}}$ and the corresponding spectral flux density
$F_{\mathrm{max}}=F\left(t_{\mathrm{max}}\right)$. Expressing then
the generic parameters $a$ and $b$ by the physical parameters $t_{\mathrm{max}}$
and $F_{\mathrm{max}}$ we find the off-axis light curve parametrization 

\begin{equation}
F_{\mathrm{off}}\!\left(t\right)=\frac{\left(\left(\frac{\beta}{\alpha}\right)^{\frac{\alpha}{\alpha+\beta}}+\left(\frac{\alpha}{\beta}\right)^{\frac{\beta}{\alpha+\beta}}\right)^{\frac{1}{\nu}}F_{\mathrm{max}}}{\left(\left(\frac{\beta}{\alpha}\right)^{\frac{\alpha}{\alpha+\beta}}\left(\frac{t}{t_{\mathrm{max}}}\right)^{-\nu\alpha}+\left(\frac{\alpha}{\beta}\right)^{\frac{\beta}{\alpha+\beta}}\left(\frac{t}{t_{\mathrm{max}}}\right)^{\nu\beta}\right)^{\frac{1}{\nu}}}\label{eq:off-axis-paremetrization}
\end{equation}
This parametrization in principle also applies to on axis events at
sufficiently low frequency, as discussed below.

\begin{figure}
\includegraphics[scale=0.3]{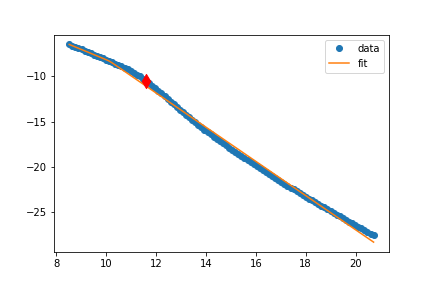}\includegraphics[scale=0.3]{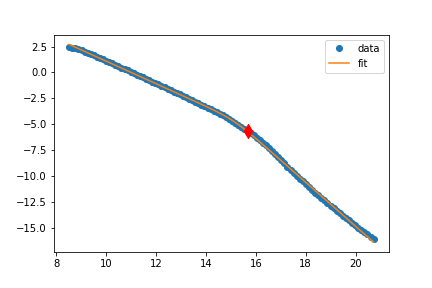}\\
\includegraphics[scale=0.3]{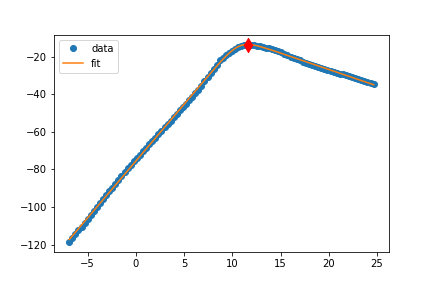}\includegraphics[scale=0.3]{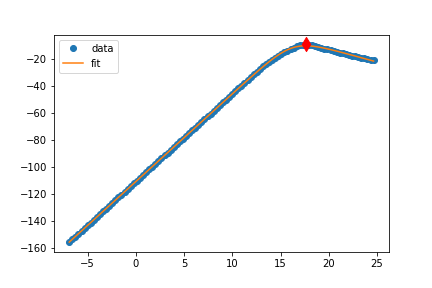}\caption{\label{fig:Fits}Fits to some exemplary infrared light curves. \textbf{Horizontal
and vertical axes show $\log\!\left(t/\mathrm{s}\right)$ and $\log\,\left(F/\mathrm{mJy}\right)$,
respectively.} On-axis light curves in the top row are fit with the
parametrization eq. (\ref{eq:on-axis-paremetrization}) and off-axis
light-curves via eq. (\ref{eq:off-axis-paremetrization}). \textbf{The
red diamond shows the position of the jet-break, determined from a
finite difference version of eq. (\ref{eq:jet-break-curvature-condition}),
respectively of the maximum.}}
\end{figure}

In the on-axis case at sufficiently high energy the light curve has
no maximum, but merely decays with different asymptotic power laws.
One could in this case define the time of jet-break as the time where
the curvature of the logarithm of the flux (which vanishes in the
asymptotic power law segments) is maximal, i.e. in terms of a logarithmic
time variable $\tau=\log t$

\begin{equation}
\frac{d^{3}}{\left(d\tau\right)^{3}}\log\left(F\left(e^{\tau_{\mathrm{jb}}}\right)\right)=0\label{eq:jet-break-curvature-condition}
\end{equation}
Unfortunately this results in very lengthy solution and therefore
does not provide a simple light curve parametrization. Therefore,
we use in the following the similar definition that the power law
exponent is just in between the two power laws (both of which give
quantitatively very similar results), i.e.

\begin{equation}
\frac{d}{d\tau}\log\left(F\left(e^{\tau_{\mathrm{jb}}}\right)\right)=\frac{\alpha-\beta}{2}
\end{equation}
This results in the even simpler parametrization

\begin{equation}
F_{\mathrm{on}}\!\left(t\right)=\frac{\sqrt[\nu]{2}F_{\mathrm{jb}}}{\left(\left(\frac{t}{t_{\mathrm{jb}}}\right)^{-\nu\alpha}+\left(\frac{t}{t_{\mathrm{jb}}}\right)^{\nu\beta}\right)^{\frac{1}{\nu}}}\label{eq:on-axis-paremetrization}
\end{equation}
The same expression is also obtained by the condition that jet break
occurs at the time where the two asymptotic power laws cross. Like
the spectral flux densities and time scales, the power law exponents
$\alpha$ and $\beta$ are physical parameters describing the asymptotic
power law behavior long before or after jet break respectively the
maximum, reflecting the effect of the jet break for an off-axis viewing
angle. However, note that while the exponent $\beta$ describes in
both cases the late-time power law decay, the parameter $\alpha$
describes, despite the same symbol, very different physical mechanisms
in the different cases. In contrast, the parameter $\nu$ does not
have a simple physical interpretation, but merely approximates the
more complicated shape of realistic light curves around jet break
respectively its maximum, see figure \ref{fig:Fits}, where the transition
region can entail more structure than our simple parametrization and
in some cases shorter third power law segments can be identified. 

In order to approximately browse the parameter space of GRB afterglow
light curves, we create a large dataset of light curves using realistic
simulations. In particular we employ the \href{https://github.com/geoffryan/afterglowpy}{AfterglowPy}
package \citep{Ryan:2019fhz} and choose a ``top-hat'' jet model.
This package is based on relativistic hydrodynamic simulations of
the forward shock in the single shell approximation \citep{vanEerten:2010zh}
and our dataset is obtained by varying key input parameters of this
model. These include the equivalent isotropic kinetic energy $E_{K,\mathrm{iso}}$,
which is determined by the central engine, as well as two environment
parameters, namely the observing angle $\theta_{\mathrm{obs}}$ and
the circumburst density $n_{0}$ of the surrounding ISM, which are
not influenced by the central engine but merely characterize the position
and orientation of the source in space. Other model parameters are
fixed to characteristic values as shown in table \ref{tab:Fixed-AfterglowPy-parameters}---the
jet opening angle, in particular, is set to $\theta_{\mathrm{jet}}=0.2$.
The main reason for this is that it is computationally hard to browse
the entire parameter space of the forward shock model, and even harder
to perform optimization in such a high-dimensional space. We vary
the kinetic energy logarithmically in the range $2\times10^{49}\,\mathrm{erg\leq}E_{K,\mathrm{iso}}\leq9\times10^{52}\,\mathrm{erg}$,
the density in the range $10^{-4}\,\mathrm{cm}^{3}\leq n\leq10^{-2}\,\mathrm{cm}^{3}$,
and the observing angle in the range $0\le\theta_{\mathrm{obs}}\leq0.18$
(on-axis) as well as $0.25\le\theta_{\mathrm{obs}}\leq0.8$ (off-axis)
to create datasets of size $16\times5\times5=400$ (on-axis) and $16\times5\times7=560$
(off-axis). Each light curve consists of 100 logarithmically-spaced
points and extends over the time interval $5\times10^{3}\,\mathrm{s}\leq t\leq5\times10^{9}\,\mathrm{s}$
in the on-axis case, respectively $10\,\mathrm{ms}\leq t\leq5\times10^{10}\,\mathrm{s}$
in the off-axis case. A few example light curves are shown in fig.
\ref{fig:Fits}.

\begin{table*}
\begin{tabular}{|c|c|}
\hline 
AfterglowPy parameter & Considered value/range\tabularnewline
\hline 
\hline 
Jet type & top-hat jet model\tabularnewline
\hline 
Jet opening angle & $\theta_{\mathrm{jet}}=0.2$ (somewhat higher than the typical value
for short GRBs which is 0.1-0.15)\tabularnewline
\hline 
Photon frequency & radio ($f=10^{10}\,\mathrm{Hz}$), near IR ($f=10^{14}\,\mathrm{Hz}$),
x-ray ($f=10^{17}\,\mathrm{Hz}$)\tabularnewline
\hline 
Source distance & $d_{L}=40\,\mathrm{Mpc}$ (distance of GRB170817)\tabularnewline
\hline 
Electron power law index & $p=2.43$ (mean value of observed short GRBs \citep{Fong:2015oha})\tabularnewline
\hline 
Magnetic energy fraction & $\epsilon_{B}=0.01$ (following \citep{Fong:2015oha})\tabularnewline
\hline 
Electron energy fraction & $\epsilon_{e}=0.1$ (following \citep{Fong:2015oha})\tabularnewline
\hline 
\hline 
Total isotropic kinetic energy & $2\times10^{49}\,\mathrm{erg\leq}E_{K,\mathrm{iso}}\leq9\times10^{52}\,\mathrm{erg}$\tabularnewline
\hline 
Observing angle & on-axis: $0\le\theta_{\mathrm{obs}}\leq0.18$ \tabularnewline
\hline 
 & off-axis: $0.25\le\theta_{\mathrm{obs}}\leq0.8$ \tabularnewline
\hline 
Circumburst density (ISM) & $10^{-4}\,\mathrm{cm}^{-3}\leq n\leq10^{-2}\,\mathrm{cm}^{-3}$\tabularnewline
\hline 
\end{tabular}.

\caption{\label{tab:Fixed-AfterglowPy-parameters}Fixed and varied AfterglowPy
parameters.}
\end{table*}
Depending on the observing angle $\theta_{\mathrm{obs}}$, we then
use either of the two parameterizations, namely eq. (\ref{eq:off-axis-paremetrization})
for $\theta_{\mathrm{obs}}>\theta_{\mathrm{jet}}$ and eq. (\ref{eq:on-axis-paremetrization})
for $\theta_{\mathrm{obs}}\leq\theta_{\mathrm{jet}}$, to fit each
light curve, using the \href{https://docs.scipy.org/doc/scipy/index.html}{SciPy}
curve-fit routine, and obtain the corresponding five light curve parameters
$F_{\mathrm{jb}}$, $t_{\mathrm{jb}}$, $\alpha$, $\beta$ and $\nu$.
Moreover, the fitting routine also provides the errors $\Delta F_{\mathrm{jb}},$...
on these quantities and we also compute the average of the corresponding
relative errors $\Delta F_{\mathrm{jb}}/F_{\mathrm{jb}}$, ... over
all light curves, as given in table \ref{tab:Averaged-fit-uncertainties},
in order to estimate the uncertainty of the fitting procedure on the
results below. The simple parameterizations eqs. (\ref{eq:off-axis-paremetrization})
and (\ref{eq:on-axis-paremetrization}) overall give a good representation
to the set of light curves, as seen for the exemplary cases in fig.
\ref{fig:Fits}. This can be quantified by the average root mean square
error, given in eq. (\ref{eq:RMSE}) below and listed in table (\ref{tab:Averaged-fit-uncertainties}),
which is at the ten percent level. As can be seen the global error
on the light curve as a whole is generally smaller than the errors
in the different parameters, which shows that the best fit chooses
slightly different values for the parameters to obtain a better global
fit. In particular, as seen in fig. \ref{fig:Fits}, the position
of jet-break can be somewhat shifted (to earlier times) in order to
make up for the fact that the synthetic light curves from the AfterglowPy
simulation have more structure than our simplified broken power law
model. This introduces a systematic error that results in the larger
uncertainties. For instance some light curves show a somewhat more
complicated structure around jet break (which could be improved by
additional short power law segments with somewhat different exponents).
Nonetheless, the parameterizations can effectively approximate such
additional structure close to jet-break via the smoothening parameter
$\nu$. Starting only at $5\times10^{3}\,\mathrm{s}$, in the on-axis
case we exclude the more uncertain plateau phase which although also
present within the forward shock model, could also stem from or be
affected by various other dynamical mechanisms. In order to correctly
fit the asymptotic power law exponents in the off-axis case we perform
the fit over the broad (and surely unobservable) time interval $10\,\mathrm{ms}\!\leq\!t\!\leq\!5\times10^{10}\,\mathrm{s}$
\footnote{Note that this is merely a computational way to prevent instabilities
in the fitting process and accurately fit the asymptotic power law
exponents---which are already clearly realized in the light curves
over realistic time intervals, see fig. \ref{fig:Fits}. In particular
we surely do not (have to) assume that the results of the simulation
are actually applicable at such extremely early and late times. In
practice the afterglow would not be visible at early times anyway
since it is completely overshadowed by the prompt emission.}

\begin{table}
\begin{tabular}{|c|c|c|c|c|c|c|}
\hline 
 &  & $\Delta F_{\mathrm{jb}}/F_{\mathrm{jb}}$ & $\Delta t_{\mathrm{jb}}/t_{\mathrm{jb}}$ & $\Delta\alpha/\alpha$ & $\Delta\beta/\beta$ & RMSE\tabularnewline
\hline 
\hline 
on-axis & x-ray & 0.428 & 0.227 & 0.066 & 0.008 & 0.151\tabularnewline
\hline 
 & IR & 0.270 & 0.143 & 0.045 & 0.006 & 0.108\tabularnewline
\hline 
 & radio & 0.075 & 0.046 & 0.108 & 0.007 & 0.088\tabularnewline
\hline 
off-axis & x-ray & 0.240 & 0.091 & 0.003 & 0.046 & 0.293\tabularnewline
\hline 
 & IR & 0.173 & 0.046 & 0.002 & 0.027 & 0.226\tabularnewline
\hline 
 & radio & 0.730 & 0.066 & 0.003 & 0.065 & 0.390\tabularnewline
\hline 
\end{tabular}

\caption{\label{tab:Averaged-fit-uncertainties}Relative uncertainties (obtained
from the standard deviations) of the corresponding parameters resulting
from the fitting procedure, averaged over all corresponding light
curves.}
\end{table}

\section{Linking in- \& output parameters of the forward shock model}

In order to obtain the leading dependencies between the three key
input parameters of the forward shock model $\left\{ \log\!\left(E_{K,\mathrm{iso}}\right),\log\!\left(\theta_{\mathrm{obs}}\right),\log\!\left(n_{0}\right)\right\} $
and the five light curve parameters $\left\{ \log\!\left(t_{\mathrm{jb}}\right),\log\!\left(F_{\mathrm{jb}}\right),\alpha,\beta,\nu\right\} $
described by the considered dataset with $400$ entries in the on-axis
case and $560$ in the off-axis case, we perform both a linear regression
using \href{https://scikit-learn.org}{Scikit-Learn} and a machine
learning analysis based on a simple neural network using the \href{https://keras.io}{Keras}
Python package. In the latter case, in order to obtain a direct dependence,
we employ a simplistic neural network with three input and five output
nodes \emph{without} intermediate hidden layers and a trivial linear
activation function. The motivation for this approach is not to obtain
the best possible model representing the data, but to train a network
whose weights manifestly encode the relation between input- and output-parameters.
The reason that we also train a neural network in this trivial case
is that this gives us an independent check of our results. Furthermore,
by generalizing the network via the inclusion of hidden layers, it
will allow us in the future to improve the analysis and consider non-linear
dependencies and this way to obtain another estimate of the validity
of the linear approximation.

The results of the linear regression and machine learning analyses
are given in terms of the slopes and offsets---in the machine learning
case called \emph{weights} and \emph{biases}---of a linear relation
between the input- ($x_{j}$) and output-parameters ($y_{i})$

\begin{equation}
y_{i}=\sum_{j}w_{ij}x_{j}+c_{i}\label{eq:linear-relation}
\end{equation}
Due to the fact that we use logarithmic variables this linear dependence
effectively amounts to a power law dependence of both the spectral
flux density and time at jet break on the kinetic energy which can
inform us about the central engine of the GRB.

\begin{table}
\begin{tabular}{|c|c|c|c|c|c||c|c|}
\hline 
$w$ &  & $\ln\!\left(\!\frac{E_{K,\mathrm{iso}}}{E_{\min}}\!\right)\!$ & $\ln\!\left(\!\frac{\theta_{\mathrm{obs}}}{\theta_{\mathrm{min}}}\!\right)\!$ & $\ln\!\left(\!\frac{n}{n_{\min}}\!\right)\!$ & $c$ & MSE & RMSE\tabularnewline
\hline 
\hline 
$\ln\!\left(\!\frac{F_{\mathrm{jb}}}{\mathrm{mJy}}\!\right)\!$ & LR & 0.940 & -0.102 & 0.925 & -12.6 & 0.0095 & 0.048\tabularnewline
\hline 
 & NN & 0.937 & -0.111 & 0.921 & -12.5 & 0.0094 & \tabularnewline
\hline 
$\ln\!\left(\!\frac{t_{\mathrm{jb}}}{\mathrm{s}}\!\right)\!$ & LR & 0.381 & 0.004 & -0.387 & 12.0 & 0.0013 & 0.017\tabularnewline
\hline 
 & NN & 0.381 & 0.004 & -0.387 & 12.0 & 0.0013 & \tabularnewline
\hline 
$\alpha$ & LR & 0.007 & -0.010 & -0.004 & -1.14 & 0.0002 & 0.014\tabularnewline
\hline 
 & NN & 0.007 & -0.010 & -0.004 & -1.14 & 0.0003 & \tabularnewline
\hline 
$\beta$ & LR & 0.019 & -0.007 & -0.019 & 1.97 & 0.00005 & 0.007\tabularnewline
\hline 
 & NN & 0.019 & -0.006 & -0.019 & 1.97 & 0.00004 & \tabularnewline
\hline 
$\nu$ & LR & -0.188 & -0.808 & 0.133 & 3.90 & 0.112 & 0.332\tabularnewline
\hline 
 & NN & -0.194 & -0.815 & 0.140 & 3.92 & 0.128 & \tabularnewline
\hline 
\end{tabular}

\caption{\label{tab:Weights-and-biases-infrared-on-axis}Weights $w_{ij}$
and biases $c_{i}$ in eq. (\ref{eq:linear-relation}), that connect
the forward shock model to the light curve parameters, for \emph{on-axis}
events at \emph{infrared} frequencies using linear regression (LR)
and neural network (NN), as well as quality metrics, namely the mean
square error (MSE) and the relative mean square error (RMSE).}
\end{table}

\begin{table}
\begin{tabular}{|c|c|c|c|c|c||c|c|}
\hline 
$w$ &  & $\ln\!\left(\!\frac{E_{K,\mathrm{iso}}}{E_{\min}}\!\right)$ & $\ln\!\left(\!\frac{\theta_{\mathrm{obs}}}{\theta_{\mathrm{min}}}\!\right)$ & $\ln\!\left(\!\frac{n}{n_{\min}}\!\right)$ & $c$ & MSE & RMSE\tabularnewline
\hline 
\hline 
$\ln\!\left(\!\frac{F_{\!\mathrm{max}}}{\mathrm{mJy}}\!\right)\!$ & LR & 1.024 & -7.960 & 0.907 & -15.8 & 0.069 & 0.129\tabularnewline
\hline 
 & NN & 1.019 & -8.007 & 0.901 & -15.8 & 0.077 & \tabularnewline
\hline 
$\ln\!\left(\!\frac{t_{\mathrm{max}}}{\mathrm{s}}\!\right)\!$ & LR & 0.339 & 4.471 & -0.380 & 12.5 & 0.019 & 0.069\tabularnewline
\hline 
 & NN & 0.342 & 4.489 & -0.377 & 12.5 & 0.019 & \tabularnewline
\hline 
$\alpha$ & LR & 0.025 & 0.206 & 0.017 & 6.16 & 0.002 & 0.041\tabularnewline
\hline 
 & NN & 0.025 & 0.207 & 0.017 & 6.16 & 0.002 & \tabularnewline
\hline 
$\beta$ & LR & 0.025 & 0.294 & -0.068 & 1.86 & 0.005 & 0.065\tabularnewline
\hline 
 & NN & 0.026 & 0.302 & -0.064 & 1.86 & 0.005 & \tabularnewline
\hline 
$\nu$ & LR & -0.012 & -0.578 & 0.069 & -2.12 & 0.006 & 0.078\tabularnewline
\hline 
 & NN & -0.029 & -0.578 & 0.069 & -2.32 & 0.009 & \tabularnewline
\hline 
\end{tabular}

\caption{\label{tab:Weights-and-biases-infrared-off-axis}Weights $w_{ij}$
and biases $c_{i}$ in eq. (\ref{eq:linear-relation}) for \emph{off-axis}
events at \emph{infrared} frequencies using linear regression (LR,
first row) and neural network (NN, second row), respectively, as well
as quality metrics, namely the mean square error (MSE) and the relative
mean square error (RMSE).}
\end{table}

The obtained values for the two computational analyses are shown for
the \emph{infrared} band in table \ref{tab:Weights-and-biases-infrared-on-axis}
for the \emph{on-axis} case and in \ref{tab:Weights-and-biases-infrared-off-axis}
for the \emph{off-axis} case. As can be seen the two computational
methods agree mostly very well with each other \footnote{This is surely not a surprise since in the considered simplified case
the neural network effectively amounts to a mere linear regression,
so that it is only the numerical optimization method that differs.}. This is also reflected in the (root) mean square error (MSE) 

\begin{equation}
\mathrm{MSE}=\sqrt{\frac{1}{n}\sum_{i=1}^{n}\left(y_{i}^{\left(\mathrm{d}\right)}-y_{i}^{\left(\mathrm{m}\right)}\right)^{2}}\label{eq:MSE}
\end{equation}
which is a measure for the deviation between the data $\left(\mathrm{d}\right)$
and the model $\left(\mathrm{m}\right)$. The linear regression and
machine learning results generally have very similar MSEs. In cases
where this is not perfectly realized, like for the smoothening parameter
in the on-axis case, this is because the actual light curve has a
more complicated functional form around jet break that our simple
power law model cannot perfectly describe, so that one of the computational
methods can have a harder time to find the absolute minimum of the
MSE in the multi-parameter space and can instead become stuck in a
local minimum.

The size of the mean square error depends on the particular light
curve parameter and cannot be directly compared. Therefore we also
consider the \emph{relative (root) mean square error} 

\begin{equation}
\mathrm{RMSE}\equiv\sqrt{\frac{1}{n}\sum_{i=1}^{n}\left(\frac{y_{i}^{\left(\mathrm{d}\right)}-y_{i}^{\left(\mathrm{m}\right)}}{y_{i}^{\left(\mathrm{d}\right)}+y_{i}^{\left(\mathrm{m}\right)}}\right)^{2}}\label{eq:RMSE}
\end{equation}
which is normalized by the size of the light curve parameter, so that
the uncertainties in the different light curve parameters can be compared.
As can be seen in all cases these relative errors are moderate which
means that the simple parameterizations eqs. (\ref{eq:off-axis-paremetrization})
and (\ref{eq:on-axis-paremetrization}) as well as the simple linear
(respectively power law) model eq. (\ref{eq:linear-relation}) can
indeed approximately describe the afterglow light curves obtained
from realistic simulations \citep{Ryan:2019fhz} and their parameter
dependencies remarkably well. Among the physical light curve parameters
it is the peak flux in the off-axis case that has the largest relative
error, which suggests that the linear model (for the logarithmic quantity)
is a bit too simplified in this case and there are sizable non-linearities
in the data that it cannot perfectly describe. Nonetheless, even in
this case the corresponding relative error is at the 10\% level which
is fine taking into account the uncertainties either in the observational
data \citep{2011A&A...528A.122P}, in key aspects of the complicated
astrophysical process \citep{Zhang:2018ond} and in its necessarily
simplified description within the hydrodynamic simulation \citep{Ryan:2019fhz}.

As discussed before, the goal of our study is to identify the imprint
of the parameter $E_{K,\mathrm{iso}}$ on the light curve morphology.
I.e. we would ideally be interested in light curve properties that
clearly change with it, but are insensitive to the two environment
parameters $\theta_{\mathrm{obs}}$ and $n$. The different weights
$w_{ij}$ depend on the sizes and ranges of the corresponding parameters
and cannot be directly compared. To compare the relative impact of
the different model parameters on the light curve parameters, we consider
the extremal contributions $\Delta y_{\mathrm{min,}ij}\equiv w_{ij}x_{j,\mathrm{min}}$
and $\Delta y_{\mathrm{max,}ij}\equiv w_{ij}x_{j,\mathrm{max}}$ over
their considered physical range of values $\left[x_{j,\mathrm{min}},x_{j,\mathrm{max}}\right]$
and define the measure

\begin{equation}
\chi_{ij}\equiv\left|\frac{\Delta y_{\mathrm{max,}ij}-\Delta y_{\mathrm{min,}ij}}{\Delta y_{\mathrm{min,}ij}+\Delta y_{\mathrm{max,}ij}}\right|/\left(\sum_{j}\left|\frac{\Delta y_{\mathrm{max,}ij}-\Delta y_{\mathrm{min,}ij}}{\Delta y_{\mathrm{min,}ij}+\Delta y_{\mathrm{max,}ij}}\right|\right)
\end{equation}
which is in the interval $\left[0,1\right]$ and quantifies how much
impact a given input parameter $x_{j}$ has on the output parameter
$y_{i}$ over the entire parameter range. It is given in table \ref{tab:Relative-impact-infrared}.
As can be seen at least over the entire considered parameter range
none of the input parameters completely dominates the others and all
are to some extent relevant for the morphology of the GRB light curve.
The only exception is that the observation angle does in the on-axis
case hardly affect the jet break position, which is expected since
we consider a top hat jet where the flux is effectively constant for
observing angles smaller than the opening angle. While in the on-axis
case the kinetic energy is the most important among the physical light
curve parameters, in the off-axis case, depending on the particular
light curve parameter, the others can be even more important.

\begin{table}
\begin{tabular}{|c|c|c|c|}
\hline 
$\chi$ & $E_{K,\mathrm{iso}}$ & $\theta_{\mathrm{obs}}$ & $n$\tabularnewline
\hline 
\hline 
$F_{\mathrm{jb}}$ & 0.64 & 0.02 & 0.34\tabularnewline
\hline 
$t_{\mathrm{jb}}$ & 0.64 & 0.002 & 0.36\tabularnewline
\hline 
$\alpha$ & 0.60 & 0.22 & 0.18\tabularnewline
\hline 
$\beta$ & 0.61 & 0.05 & 0.34\tabularnewline
\hline 
$\nu$ & 0.39 & 0.46 & 0.15\tabularnewline
\hline 
\end{tabular}$\qquad$%
\begin{tabular}{|c|c|c|c|}
\hline 
$\chi$ & $E_{K,\mathrm{iso}}$ & $\theta_{\mathrm{obs}}$ & $n$\tabularnewline
\hline 
\hline 
$F_{\mathrm{jb}}$ & 0.43 & 0.37 & 0.20\tabularnewline
\hline 
$t_{\mathrm{jb}}$ & 0.33 & 0.47 & 0.20\tabularnewline
\hline 
$\alpha$ & 0.44 & 0.40 & 0.16\tabularnewline
\hline 
$\beta$ & 0.28 & 0.35 & 0.37\tabularnewline
\hline 
$\nu$ & 0.11 & 0.56 & 0.33\tabularnewline
\hline 
\end{tabular}

\caption{\label{tab:Relative-impact-infrared}Relative impact $\chi_{ij}$
of the different model input parameters in eq. (\ref{eq:linear-relation})
for \emph{on-axis} (left table) and \emph{off-axis} events (right
table) at \emph{infrared} frequencies.}
\end{table}

Due to the linear relation eq. (\ref{eq:linear-relation}) between
the quantities, the logarithmic variables obey simple power law relations.
As an example, the spectral flux density at jet-break, obtained from
the linear regression (respectively neural network) results in the
on-axis case in table \ref{tab:Weights-and-biases-infrared-on-axis},
takes the explicit form

\begin{align}
F_{\mathrm{jb}}^{\left(\mathrm{IR,on}\right)} & =4.8\times10^{5}\,\mu\mathrm{Jy}\left(\frac{E_{K,\mathrm{iso}}}{10^{52}\,\mathrm{erg}}\right)^{0.94}\left(\frac{\theta_{\mathrm{obs}}}{\theta_{0}}\right)^{-0.10}\nonumber \\
 & \qquad\qquad\times\left(\frac{n}{\mathrm{cm}^{-3}}\right)^{0.93}\left(\frac{d_{L}}{40\,\mathrm{Mpc}}\right)^{-2}\label{eq:IR-flux}
\end{align}
which taking into account the redshifting gives the flux at an IR
frequency $\zeta\!\left(z\right)\times10^{10}\,\mathrm{Hz}$, where
the relative redshifting factor is

\begin{equation}
\zeta\!\left(z\right)\equiv\frac{1+z}{1+40\,\mathrm{Mpc}H_{0}/c}\approx0.9999\left(1+z\right)\approx1+z
\end{equation}
Eq. (\ref{eq:IR-flux}) depends with a similar power on the kinetic
energy and the density, but hardly depends on the angle (varying over
a narrow interval), as already seen in table \ref{tab:Relative-impact-infrared}.

\begin{table}
on-axis:

\begin{tabular}{|c|c|c|c|c||c|c|}
\hline 
$w$ & $\ln\!\left(\!\frac{E_{K,\mathrm{iso}}}{E_{\min}}\!\right)$ & $\ln\!\left(\!\frac{\theta_{\mathrm{obs}}}{\theta_{\mathrm{min}}}\!\right)$ & $\ln\!\left(\!\frac{n}{n_{\min}}\!\right)$ & $c$ & MSE & RMSE\tabularnewline
\hline 
\hline 
$\ln\!\left(\frac{F_{\mathrm{jb}}}{\mathrm{mJy}}\right)$ & 0.9952 & 0.034 & 1.007 & -13.81 & 0.039 & 0.098\tabularnewline
\hline 
$\ln\!\left(\frac{t_{\mathrm{jb}}}{\mathrm{s}}\right)$ & 0.2790 & -0.047 & -0.531 & 12.46 & 0.051 & 0.112\tabularnewline
\hline 
$\alpha$ & -0.0002 & -0.001 & -0.005 & -1.12 & 60520 & 0.008\tabularnewline
\hline 
$\beta$ & 0.0154 & -0.006 & -0.013 & 1.89 & 0.030 & 0.005\tabularnewline
\hline 
$\nu$ & -0.2858 & -0.333 & -0.098 & 4.38 & 0.342 & 0.746\tabularnewline
\hline 
\end{tabular}

off-axis:

\begin{tabular}{|c|c|c|c|c||c|c|}
\hline 
$w$ & $\ln\!\left(\!\frac{E_{K,\mathrm{iso}}}{E_{\min}}\!\right)$ & $\ln\!\left(\!\frac{\theta_{\mathrm{obs}}}{\theta_{\mathrm{min}}}\!\right)$ & $\ln\!\left(\!\frac{n}{n_{\min}}\!\right)$ & $c$ & MSE & RMSE\tabularnewline
\hline 
\hline 
$\ln\!\left(\!\frac{F_{\mathrm{max}}}{\mathrm{mJy}}\!\right)$ & 0.851 & -8.108 & 0.658 & -20.5 & 0.165 & 0.196\tabularnewline
\hline 
$\ln\!\left(\!\frac{t_{\mathrm{max}}}{\mathrm{s}}\!\right)$ & 0.373 & 4.948 & -0.352 & 12.4 & 0.022 & 0.073\tabularnewline
\hline 
$\alpha$ & 0.002 & 0.005 & -0.005 & 6.50 & 0.0001 & 0.009\tabularnewline
\hline 
$\beta$ & 0.078 & 0.826 & -0.110 & 1.77 & 0.0337 & 0.171\tabularnewline
\hline 
$\nu$ & -0.021 & -0.558 & 0.061 & -2.50 & 0.0019 & 0.044\tabularnewline
\hline 
\end{tabular}

\caption{\label{tab:Weights-and-biases-x-ray}Weights $w_{ij}$ and biases
$c_{i}$ in eq. (\ref{eq:linear-relation}) at \emph{x-ray} frequencies,
as well as absolute (MSE) and relative mean square error (RMSE).}
\end{table}

\begin{table}
on-axis:

\begin{tabular}{|c|c|c|c|c||c|c|}
\hline 
$w$ & $\ln\!\left(\!\frac{E_{K,\mathrm{iso}}}{E_{\min}}\!\right)$ & $\ln\!\left(\!\frac{\theta_{\mathrm{obs}}}{\theta_{\mathrm{min}}}\!\right)$ & $\ln\!\left(\!\frac{n}{n_{\min}}\!\right)$ & $c$ & MSE & RMSE\tabularnewline
\hline 
\hline 
$\ln\!\left(\frac{F_{\mathrm{jb}}}{\mathrm{mJy}}\right)$ & 0.987 & -0.059 & 0.537 & -1.80 & 0.002 & 0.021\tabularnewline
\hline 
$\ln\!\left(\frac{t_{\mathrm{jb}}}{\mathrm{s}}\right)$ & 0.352 & -0.016 & -0.165 & 11.96 & 0.0003 & 0.009\tabularnewline
\hline 
$\alpha$ & 0.009 & 0.015 & 0.002 & 0.42 & 0.0003 & 0.017\tabularnewline
\hline 
$\beta$ & 0.023 & -0.004 & -0.029 & 2.00 & 0.067 & 0.008\tabularnewline
\hline 
$\nu$ & -0.075 & -0.192 & 0.212 & 0.40 & 0.011 & 0.107\tabularnewline
\hline 
\end{tabular}

off-axis:

\begin{tabular}{|c|c|c|c|c||c|c|}
\hline 
$w$ & $\ln\!\left(\!\frac{E_{K,\mathrm{iso}}}{E_{\min}}\!\right)$ & $\ln\!\left(\!\frac{\theta_{\mathrm{obs}}}{\theta_{\mathrm{min}}}\!\right)$ & $\ln\!\left(\!\frac{n}{n_{\min}}\!\right)$ & $c$ & MSE & RMSE\tabularnewline
\hline 
\hline 
$\ln\!\left(\!\frac{F_{\mathrm{max}}}{\mathrm{mJy}}\!\right)$ & 0.999 & -6.201 & 0.795 & -4.67 & 0.017 & 0.065\tabularnewline
\hline 
$\ln\!\left(\!\frac{t_{\mathrm{max}}}{\mathrm{s}}\!\right)$ & 0.330 & 3.385 & -0.310 & 13.31 & 0.0044 & 0.033\tabularnewline
\hline 
$\alpha$ & 0.020 & 0.173 & 0.009 & 5.75 & 0.0009 & 0.030\tabularnewline
\hline 
$\beta$ & 0.015 & -0.023 & -0.011 & 1.81 & 0.0001 & 0.012\tabularnewline
\hline 
$\nu$ & -0.115 & -0.059 & 0.018 & 0.16 & 0.062 & 0.250\tabularnewline
\hline 
\end{tabular}

\caption{\label{tab:Weights-and-biases-radio}Weights $w_{ij}$ and biases
$c_{i}$ in eq. (\ref{eq:linear-relation}) at \emph{radio} frequencies,
as well as absolute (MSE) and relative mean square error (RMSE).}
\end{table}
In addition to the infrared light curves discussed so far we have
performed the same analysis also for the x-ray and radio band and
the results are given in tables \ref{tab:Weights-and-biases-x-ray}
and \ref{tab:Weights-and-biases-radio}. As in the infrared case the
model parametrization can also in this case reproduce the AfterglowPy
results well and provide a simple parametrization of the physical
light curve parameters in terms of forward shock model parameters
valid at the tens of percent level. Merely the smoothing parameter
$\nu$ is not well constrained in the on-axis case at x-ray frequencies
signalizing that there is more structure around jet-break in these
light curves than our simple model can capture. Moreover, in this
case the plateau phase might still have some impact on the early time
behavior. As is well known the key difference of on-axis light curves
in the radio band is that the light curve is initially increasing
\citep{Zhang:2018ond} (while they are monotonously decreasing at
higher frequencies), because the shock emission is initially very
energetic and only over time moves towards lower energies/frequencies
as the shock slows down and/or the injection from the central engine
weakens. Since at the same time radio light curves do not show a pronounced
jet break, the parameter $\alpha$ fits in this case the power law
rise \footnote{It might seem more natural to rather employ the off-axis parameterization
eq. (\ref{eq:off-axis-paremetrization}) at radio frequencies where
the light curve has a maximum (so that the spectral flux density and
time reflect the actual values at this maximum), but to perform a
consistent frequency dependent multi-linear regression below, we refrain
from this at this point.}.

\begin{table*}
on-axis:

\begin{tabular}{|c|c|c|c|c|c||c|c|}
\hline 
$w$ & $\ln\!\left(\frac{E_{K,\mathrm{iso}}}{E_{\min}}\right)$ & $\ln\!\left(\frac{\theta_{\mathrm{obs}}}{\theta_{\mathrm{min}}}\right)$ & $\ln\!\left(\frac{n}{n_{\min}}\right)$ & $\ln\!\left(\frac{f}{f_{\min}}\right)$ & $c$ & MSE & RMSE\tabularnewline
\hline 
\hline 
$\ln\!\left(\frac{F_{\mathrm{max}}}{\mathrm{mJy}}\right)$ & 0.971 & -0.067 & 0.803 & -0.672 & -3.36 & 3.053 & 0.657\tabularnewline
\hline 
$\ln\!\left(\frac{t_{\mathrm{max}}}{\mathrm{s}}\right)$ & 0.338 & -0.022 & -0.359 & -0.043 & 12.52 & 0.106 & 0.158\tabularnewline
\hline 
$\alpha$ & 0.005 & -0.004 & -0.006 & -0.102 & 0.31 & 0.090 & 0.148\tabularnewline
\hline 
$\beta$ & 0.019 & -0.005 & -0.020 & -0.007 & 2.01 & 0.001 & 0.011\tabularnewline
\hline 
$\nu$ & -1.929 & -5.307 & 0.681 & 0.955 & 16.74 & 62.00 & 0.892\tabularnewline
\hline 
\end{tabular}

off-axis:

\begin{tabular}{|c|c|c|c|c|c||c|c|}
\hline 
$w$ & $\ln\!\left(\frac{E_{K,\mathrm{iso}}}{E_{\min}}\right)$ & $\ln\!\left(\frac{\theta_{\mathrm{obs}}}{\theta_{\mathrm{min}}}\right)$ & $\ln\!\left(\frac{n}{n_{\min}}\right)$ & $\ln\!\left(\frac{f}{f_{\min}}\right)$ & $c$ & MSE & RMSE\tabularnewline
\hline 
\hline 
$\ln\!\left(\frac{F_{\mathrm{max}}}{\mathrm{mJy}}\right)$ & 0.955 & -8.398 & 0.762 & -0.863 & -3.69 & 0.695 & 0.380\tabularnewline
\hline 
$\ln\!\left(\frac{t_{\mathrm{max}}}{\mathrm{s}}\right)$ & 0.354 & 3.862 & -0.342 & 0.0066 & 12.70 & 0.601 & 0.334\tabularnewline
\hline 
$\alpha$ & 0.022 & 0.207 & 0.011 & 0.033 & 5.72 & 0.007 & 0.043\tabularnewline
\hline 
$\beta$ & 0.040 & 0.259 & -0.056 & 0.023 & 1.63 & 0.044 & 0.101\tabularnewline
\hline 
$\nu$ & -0.033 & -0.159 & 0.011 & -0.045 & 0.93 & 0.040 & 0.096\tabularnewline
\hline 
\end{tabular}

\caption{\label{tab:Weights-and-biases-broadband}Weights $w_{ij}$ and biases
$c_{i}$, for broadband afterglows as well as quality metrics.}
\end{table*}

Finally, combining these three data sets (x-ray, infrared and radio
with 1200 on-axis and 1680 off-axis light curves overall), we use
the frequency as another free parameter for the multi-linear regression
analysis in order to check whether even a general parameterization
of the combined AfterglowPy light curves is possible, that is valid
at any frequency, and the results are shown in table \ref{tab:Weights-and-biases-broadband}.
As can be seen our simple linear/power law model is in general too
simplified to obtain an accurate representation of the short GRB light
curves over the entire electromagnetic spectrum. The relative uncertainties
are in this case typically in the many tens of percent range even
for the physical light curve parameters (and not just for the unphysical
smoothing parameter $\nu$). The main reason is the pronounced frequency
dependence of the qualitative shape of the on-axis light curve, mentioned
above, where the power law exponent $\alpha$ that is standardly negative
in the normal decay phase and not too different at x-ray and infrared
frequencies (see the parameter offset parameter $c$ in table \ref{tab:Weights-and-biases-infrared-on-axis}
and \ref{tab:Weights-and-biases-x-ray}), changes sign due to the
rise in the radio data (see table \ref{tab:Weights-and-biases-radio}),
introducing a pronounced non-linearity that naturally cannot be captured
by our linear regression model. Despite these shortcomings the results
might be useful nonetheless as a rough estimate of light curves in
other bands of the electromagnetic spectrum. As an example of a quantity
that is reasonably well constrained, the time where the jet-break
in the on-axis light curve occurs takes the general form 

\begin{align}
t_{\mathrm{jb}}^{\left(\mathrm{on}\right)} & \approx\left(1.6\pm0.3\right)\mathrm{days}\left(\frac{E_{K,\mathrm{iso}}}{10^{50}\,\mathrm{erg}}\right)^{0.34}\left(\frac{\theta_{\mathrm{obs}}}{0.25}\right)^{-0.02}\nonumber \\
 & \qquad\times\left(\frac{n}{10^{-3}\,\mathrm{cm}^{-3}}\right)^{-0.36}\left(\frac{f}{\zeta\times10^{14}\,\mathrm{Hz}}\right)^{-0.04}
\end{align}
showing that for a top hat jet shape the light curve does effectively
not depend on the observing angle, as it should. It also shows that
the forward shock mechanism is effectively achromatic in accordance
with most observed afterglow light curves. This includes in particular
the radio band showing that the maximum in the light curve is connected
to the jet-break. The (slightly) negative exponent is also in line
with some observed chromatic light curves where lower energy radiation
indeed shows a jet break at a later time \citep{Zhang:2018ond}. Finally,
it is interesting to note that the energy and density dependence is
close to the theoretical expectation $\sim\left(E_{k,\mathrm{iso}}/n\right)^{1/3}$
\citep{Zhang:2009tr}.

The late-time power law decay is even more accurately determined and
hardly changes throughout the browsed parameter space, so that it
is constrained over the entire afterglow spectrum to the narrow interval

\begin{equation}
\beta^{\left(\mathrm{on}\right)}\approx2.01_{-0.21}^{+0.16}
\end{equation}
In contrast, due to this large power and the uncertainty on the jet-break
time, the maximum spectral flux density at jet-break naturally has
an even lager uncertainty. The same holds for the parameter $\nu$,
which is likewise expected taking into account the very different
physics responsible for the early time behavior of these light curves
in the different bands. 

In the off-axis case in the considered top-hat jet model, the parameter
$\alpha$, describing the power law rise before the maximum, is only
weekly dependent on the forward shock parameters and depends most
sensitively on the observing angle (which has a rather narrow range,
though)

\begin{align}
 & \alpha^{\left(\mathrm{off}\right)}\!\approx\!5.72\!\pm\!0.07\!+\!0.02\log\!\left(\!\frac{E_{K,\mathrm{iso}}}{2\times10^{49}\,\mathrm{erg}}\!\right)\!+\!0.21\log\!\left(\!\frac{\theta_{\mathrm{obs}}}{0.25}\!\right)\nonumber \\
 & \quad+\!0.01\log\!\left(\frac{n}{10^{-3}\,\mathrm{cm}^{-3}}\right)\!+\!0.03\log\!\left(\frac{f}{10^{14}\,\mathrm{Hz}}\right)
\end{align}
I.e. a top-hat jet considered here yields a rapidly rising flux, while
structured jets typically feature a much more shallow rise, as seen
for GRB 170717A \citep{LIGOScientific:2017ync}.

\section{Strength of the central engine from afterglow light curves}

Having found that the output of the forward shock model obtained from
the AfterglowPy simulations can be reasonably well approximated by
the simple broken power law models, we now solve the inverse problem
in order to obtain the key parameter that is determined by the central
engine, namely the unknown equivalent isotropic kinetic energy $E_{K,\mathrm{iso}}$,
from observed short GRB light curves, described by the corresponding
light curve parameters. To this end we use the same data set and simply
perform the linear regression the other way around. For on-axis radio
light curves, that have a maximum, since the typical synchrotron energies
decrease over time and only eventually reach the radio band, we use
the parametrization eq. (\ref{eq:off-axis-paremetrization}) as in
the off-axis case. In terms of the five light curve parameters the
results are shown in the first rows of tables \ref{tab:Weights-and-biases-energy-x-ray}
to \ref{tab:Weights-and-biases-energy-radio} at x-ray, infrared and
radio frequencies, respectively. As can be seen the kinetic energy
can be estimated with an average uncertainty in the percent range
when all five light curve parameters can be determined.

\begin{table}
on-axis:

\begin{tabular}{|c|c|c|c|c|c||c|}
\hline 
$\ln\!\left(\frac{F_{\mathrm{jb}}}{\mathrm{mJy}}\right)$ & $\ln\!\left(\frac{t_{\mathrm{jb}}}{\mathrm{s}}\right)$ & $\beta$ & $\alpha$ & $\nu$ & $c$ & RMSE\tabularnewline
\hline 
\hline 
0.565 & 0.941 & 9.250 & -10.15 & -7474 & -32.37 & 0.077\tabularnewline
\hline 
0.578 & 1.054 & 7.844 & -16.67 & --- & -38.49 & 0.081\tabularnewline
\hline 
0.581 & 0.773 & 12.83 & --- & --- & -25.66 & 0.103\tabularnewline
\hline 
0.689 & 1.120 & --- & --- & --- & -4.625 & 0.181\tabularnewline
\hline 
0.707 & --- & --- & --- & --- & 9.273 & 0.541\tabularnewline
\hline 
\end{tabular}

off-axis:

\begin{tabular}{|c|c|c|c|c|c||c|}
\hline 
$\ln\!\left(\frac{F_{\mathrm{jb}}}{\mathrm{mJy}}\right)$ & $\ln\!\left(\frac{t_{\mathrm{jb}}}{\mathrm{s}}\right)$ & $\beta$ & $\alpha$ & $\nu$ & $c$ & RMSE\tabularnewline
\hline 
\hline 
0.637 & 1.352 & -1.024 & 28.83 & 22.60 & -193.6 & 0.078\tabularnewline
\hline 
0.668 & 1.239 & -1.048 & 24.08 & --- & -159.0 & 0.108\tabularnewline
\hline 
0.666 & 1.087 & -0.274 & --- & --- & -3.265 & 0.110\tabularnewline
\hline 
0.722 & 1.165 & --- & --- & --- & -3.534 & 0.140\tabularnewline
\hline 
0.473 & --- & --- & --- & --- & 11,67 & 0.638\tabularnewline
\hline 
\end{tabular}

\caption{\label{tab:Weights-and-biases-energy-x-ray} Estimates from \emph{x-ray}
light curves for the kinetic energy $\ln\!\left(\!E_{K,\mathrm{iso}}/E_{\min}\!\right)$
within the forward shock model, characterizing the strength of the
central engine, as well as the relative mean square error (RMSE).}
\end{table}

\begin{table}

on-axis:

\begin{tabular}{|c|c|c|c|c|c||c|}
\hline 
$\ln\!\left(\frac{F_{\mathrm{jb}}}{\mathrm{mJy}}\right)$ & $\ln\!\left(\frac{t_{\mathrm{jb}}}{\mathrm{s}}\right)$ & $\beta$ & $\alpha$ & $\nu$ & $c$ & RMSE\tabularnewline
\hline 
\hline 
0.538 & 1.321 & -1.074 & 0.452 & -0.003 & -6.289 & 0.017\tabularnewline
\hline 
0.540 & 1.454 & -3.250 & -0.082 & --- & -4.255 & 0.023\tabularnewline
\hline 
0.540 & 1.457 & -3.353 & --- & --- & -4.020 & 0.023\tabularnewline
\hline 
0.626 & 1.102 & --- & --- & --- & -5.700 & 0.156\tabularnewline
\hline 
0.765 & --- & --- & --- & --- & 9.213 & 0.543\tabularnewline
\hline 
\end{tabular}

off-axis:

\begin{tabular}{|c|c|c|c|c|c||c|}
\hline 
$\ln\!\left(\frac{F_{\mathrm{jb}}}{\mathrm{mJy}}\right)$ & $\ln\!\left(\frac{t_{\mathrm{jb}}}{\mathrm{s}}\right)$ & $\beta$ & $\alpha$ & $\nu$ & $c$ & RMSE\tabularnewline
\hline 
\hline 
0.630 & 1.076 & 0.765 & -1.504 & -5.630 & 4.815 & 0.072\tabularnewline
\hline 
0.615 & 1.157 & 0.624 & -1.111 & --- & 0.510 & 0.079\tabularnewline
\hline 
0.598 & 1.044 & 0.867 & --- & --- & -5.483 & 0.108\tabularnewline
\hline 
0.628 & 1.027 & --- & --- & --- & -3.631 & 0.188\tabularnewline
\hline 
0.313 & --- & --- & --- & --- & 8.374 & 0.664\tabularnewline
\hline 
\end{tabular}

\caption{\label{tab:Weights-and-biases-energy-inrared} Estimates from \emph{infrared}
light curves for the kinetic energy $\ln\!\left(\!E_{K,\mathrm{iso}}/E_{\min}\!\right)$
within the forward shock model, characterizing the strength of the
central engine, as well as the relative mean square error (RMSE).}
\end{table}

\begin{table}[t]

on-axis:

\begin{tabular}{|c|c|c|c|c|c||c|}
\hline 
$\ln\!\left(\frac{F_{\mathrm{jb}}}{\mathrm{mJy}}\right)$ & $\ln\!\left(\frac{t_{\mathrm{jb}}}{\mathrm{s}}\right)$ & $\beta$ & $\alpha$ & $\nu$ & $c$ & RMSE\tabularnewline
\hline 
\hline 
0.564 & 0.766 & 6.245 & 2.619 & -0.166 & -21.19 & 0.024\tabularnewline
\hline 
0.529 & 0.810 & 7.360 & 3.888 & --- & -24.75 & 0.043\tabularnewline
\hline 
0.676 & 0.199 & 11.082 & --- & --- & -23.26 & 0.059\tabularnewline
\hline 
0.176 & 1.591 & --- & --- & --- & -17.28 & 0.145\tabularnewline
\hline 
0.907 & --- & --- & --- & --- & 0.824 & 0.365\tabularnewline
\hline 
\end{tabular}

off-axis:

\begin{tabular}{|c|c|c|c|c|c||c|}
\hline 
$\ln\!\left(\frac{F_{\mathrm{jb}}}{\mathrm{mJy}}\right)$ & $\ln\!\left(\frac{t_{\mathrm{jb}}}{\mathrm{s}}\right)$ & $\beta$ & $\alpha$ & $\nu$ & $c$ & RMSE\tabularnewline
\hline 
\hline 
0.549 & 1.106 & 6.101 & -1.046 & -0.136 & -17.07 & 0.041\tabularnewline
\hline 
0.552 & 1.112 & 6.467 & -0.982 & --- & -18.32 & 0.049\tabularnewline
\hline 
0.538 & 1.050 & 7.385 & --- & --- & -24.89 & 0.075\tabularnewline
\hline 
0.411 & 1.660 & --- & --- & --- & -18.61 & 0.235\tabularnewline
\hline 
0.565 & --- & --- & --- & --- & 5.052 & 0.588\tabularnewline
\hline 
\end{tabular}

\caption{\label{tab:Weights-and-biases-energy-radio} Estimates from \emph{radio}
light curves for the kinetic energy $\ln\!\left(\!E_{K,\mathrm{iso}}/E_{\min}\!\right)$
within the forward shock model, characterizing the strength of the
central engine, as well as the relative mean square error (RMSE).}
\end{table}

Here we are in particular interested how much information about the
light curve is required to obtain a reliable estimate for $E_{K,\mathrm{iso}}$.
In order to test whether fewer parameters are sufficient, we have
performed linear regression for the kinetic energy as a function of
different sets of these parameters ranging from all five light curve
parameters down to a single light curve parameter.

As discussed before, the smoothing parameter $\nu$ is an unphysical
fitting parameter that merely takes into account that the light curve
can have more structure around jet break in the on-axis case, respectively
around the maximum in the off-axis case. Nonetheless it is required
to obtain the correct values for the other light curve parameters
$F_{\mathrm{\mathrm{max/jb}}}$, $t_{\mathrm{max/jb}}$ and in particular
the asymptotic power law exponents $\alpha$ and $\beta$ that tell
us something about the physical process. Moreover, $\nu$ inherently
cannot simply be read off a given light curve and therefore it would
surely be advantageous if it is not needed to predict $E_{K,\mathrm{iso}}$.
As can be seen in the second row of tables \ref{tab:Weights-and-biases-energy-x-ray}
to \ref{tab:Weights-and-biases-energy-radio}, only the knowledge
of the four physical parameters yields the kinetic energy with a similar
average relative error. Using these value, the resulting estimate
for the kinetic energy \emph{in terms of the four physical light curve
parameters} reads then explicitly in the different cases

\begin{widetext}

\begin{align}
E_{K,\mathrm{iso}}^{\mathrm{\left(on\right)}} & =\begin{cases}
\left(9.8\pm1.6\pm16.5\right)\times10^{49}\,\mathrm{erg}\left(\frac{F_{\mathrm{jb}}}{\mathrm{nJy}}\right)^{0.58}\left(\frac{t_{\mathrm{jb}}}{\mathrm{day}}\right)^{1.05}e^{7.8\left(\beta-2\right)}e^{-16.7\left(\alpha+1\right)}\left(\frac{d_{L}}{\mathrm{Gpc}}\right)^{1.16} & \mathrm{x-ray}\:\left(\zeta\times10^{17}\,\mathrm{Hz}\right)\\
\left(1.3\pm0.1\pm0.5\right)\times10^{50}\,\mathrm{erg}\left(\frac{F_{\mathrm{jb}}}{\mathrm{nJy}}\right)^{0.54}\left(\frac{t_{\mathrm{jb}}}{\mathrm{day}}\right)^{1.45}e^{-3.3\left(\beta-2\right)}e^{-0.1\left(\alpha+1\right)}\left(\frac{d_{L}}{\mathrm{Gpc}}\right)^{1.08} & \mathrm{infrared}\:\left(\zeta\times10^{14}\,\mathrm{Hz}\right)\\
\left(3.4\pm0.3\pm1.4\right)\times10^{50}\,\mathrm{erg}\left(\frac{F_{\mathrm{jb/max}}}{\mathrm{\mu Jy}}\right)^{0.53}\left(\frac{t_{\mathrm{jb/max}}}{\mathrm{day}}\right)^{0.81}e^{7.36\left(\beta-2\right)}e^{3.89\left(\alpha-0.5\right)}\left(\frac{d_{L}}{\mathrm{Gpc}}\right)^{1.06} & \mathrm{radio}\:\left(\zeta\times10^{10}\,\mathrm{Hz}\right)
\end{cases}\label{eq:E-on-4-parameters}
\end{align}

\begin{align}
E_{K,\mathrm{iso}}^{\mathrm{\left(off\right)}} & =\begin{cases}
\left(1.9\pm0.7\pm0.8\right)\times10^{51}\,\mathrm{erg}\left(\frac{F_{\mathrm{max}}}{\mathrm{nJy}}\right)^{0.67}\left(\frac{t_{\mathrm{max}}}{\mathrm{day}}\right)^{1.24}e^{-1.05\left(\beta-2\right)}e^{24.1\left(\alpha-6.5\right)}\left(\frac{d_{L}}{\mathrm{Gpc}}\right)^{1.34} & \mathrm{x-ray}\:\left(\zeta\times10^{17}\,\mathrm{Hz}\right)\\
\left(4.7\pm0.7\pm0.8\right)\times10^{50}\,\mathrm{erg}\left(\frac{F_{\mathrm{max}}}{\mathrm{nJy}}\right)^{0.62}\left(\frac{t_{\mathrm{max}}}{\mathrm{day}}\right)^{1.16}e^{0.62\left(\beta-2\right)}e^{-1.11\left(\alpha-6.5\right)}\left(\frac{d_{L}}{\mathrm{Gpc}}\right)^{1.23} & \mathrm{infrared}\:\left(\zeta\times10^{14}\,\mathrm{Hz}\right)\\
\left(3.7\pm0.4\pm3.3\right)\times10^{49}\,\mathrm{erg}\left(\frac{F_{\mathrm{max}}}{\mathrm{\mu Jy}}\right)^{0.55}\left(\frac{t_{\mathrm{max}}}{\mathrm{day}}\right)^{1.11}e^{6.47\left(\beta-2\right)}e^{-0.98\left(\alpha-6.5\right)}\left(\frac{d_{L}}{\mathrm{Gpc}}\right)^{1.10} & \mathrm{radio}\:\left(\zeta\times10^{10}\,\mathrm{Hz}\right)
\end{cases}\label{eq:E-off-4-parameters}
\end{align}

\end{widetext}

To estimate the uncertainties, we use twice the RMSE value as an error
estimate for the linear regression (first given error) and employ
linear error propagation to estimate the error stemming from the light
curve fitting procedure in all light curve parameters (second error).
Adding all these errors will on the one hand likely significantly
overestimate the actual errors due to our simplified parameterization
and numerical analysis. \footnote{This is apparent since using linear error propagation, the errors
in the 4-parameter estimate in eqs. (\ref{eq:E-on-4-parameters})
and (\ref{eq:E-off-4-parameters}) are formally significantly larger
than the one in the 2-parameter result given below in eqs. (\ref{eq:E-on-2-parameters})
and (\ref{eq:E-off-2-parameters}).} On the other hand these errors do not include systematic uncertainties
due to approximations within the forward shock model or the considered
top-hat jet shape---in the on-axis case the latter should be a very
good approximation, though. In these expressions we also added the
trivial (luminosity) distance dependence, and we give them in terms
of distance instead of redshift since the distance-redshift relation
does not have a simple form at the large redshifts $z\lesssim1$ of
typical short GRBs.

Also the exponent $\alpha$, determining the power law pre jet break
respectively the maximum, might not be known for a given light curve
since the afterglow has not been observed early enough to obtain it.
Another reason is that the early time behavior in real light curves
can be overshadowed by the signal from the reverse shock \citep{Zhang:2018ond},
so that it is hard to distinguish the forward and reverse shock signal
and determine $\alpha$. In such cases it would be advantageous if
the kinetic energy could be reliably determined with the information
of only the late time parameters of the light curve. The results are
given in row three of tables \ref{tab:Weights-and-biases-energy-x-ray}
to \ref{tab:Weights-and-biases-energy-radio} showing the result in
terms only the three late time parameters $F_{\mathrm{max/jb}}$,
$t_{\mathrm{max/jb}}$ and $\beta$. Although slightly less precise,
the late time parameters are sufficient to determine the energy, still
with a comparable average error. This is very fortunate since it circumvents
the adverse impact of dynamical processes following the prompt emission,
in addition to the likely strongly magnetized reverse shock or even
flares, also jet features like its structure or a surrounding cocoon,
that are all not included in the present top-hat forward shock model.
The late time behavior in contrast should be reasonably well described
by the emission from the forward shock. And in the on-axis case, which
is realized for most of the observed GRB afterglows, our result based
on a top hat geometry should be well applicable and provide reliable
information on the strength of the central engine. 

As can be seen in tables \ref{tab:Weights-and-biases-energy-x-ray}
to \ref{tab:Weights-and-biases-energy-radio} as fewer parameters
are considered the power law exponents of the coefficients in some
cases only very mildly change, which suggests that the omitted parameters
do not have a big impact on $E_{K,\mathrm{iso}}$. In other cases
the remaining parameters change significantly and effectively compensate
for the impact of neglected parameters to still provide a reasonable
relative error. That this works suggests that the dependencies on
these light curve parameters are correlated and fewer of them are
enough to obtain reliable information on the strength of the central
engine. I.e. the dependencies on the remaining power law exponents
themselves are not faithfully approximated in these cases but the
obtained \emph{effective} power law on average nonetheless estimates
the kinetic energy with sufficient accuracy.

Similarly the exponent $\beta$, determining the late time decay,
might not be known for an observed afterglow since the afterglow is
already too faint to observe it long enough with a given observatory,
in order to clearly determine the decay exponent. Moreover, this parameter
can only be reliable determined at late time and is therefore inherently
unknown initially. Row four of tables \ref{tab:Weights-and-biases-energy-x-ray}
to \ref{tab:Weights-and-biases-energy-radio} shows the results when
only the position of the jet break or maximum, given by $F_{\mathrm{max/jb}}$
and $t_{\mathrm{max/jb}}$, is used. In this case it is not even necessary
to perform fitting to determine these light curve parameters, but
they are e.g. precisely determined by the criterium that at jet break
the curvature is maximal, eq. \ref{eq:jet-break-curvature-condition}.
The curvature can be easily determined by finite differencing, so
that the error is here smaller, merely determined by the grid spacing
of the synthetic light curve table (second error), \footnote{We use for the error on the position of both the jet-break and the
maximum the values half a grid spacing away in either direction and
average over all light curves. Increasing the size of temporal grid
points this discretization error.} and could in principle easily be decreased. The expression for the
kinetic energy takes, \emph{in terms of the two parameters characterizing
the position of the jet-break respectively maximum,} the explicit
form in the different cases

\begin{widetext}

\begin{align}
E_{K,\mathrm{iso}}^{\mathrm{\left(on\right)}} & =\begin{cases}
\left(4.1\pm1.5\pm0.7\right)\times10^{50}\,\mathrm{erg}\left(\frac{F_{\mathrm{jb}}}{\mathrm{nJy}}\right)^{0.69}\left(\frac{t_{\mathrm{jb}}}{\mathrm{day}}\right)^{1.12}\left(\frac{d_{L}}{\mathrm{Gpc}}\right)^{1.38} & \mathrm{x-ray}\:\left(\zeta\times10^{17}\,\mathrm{Hz}\right)\\
\left(1.8\pm0.6\pm0.3\right)\times10^{50}\,\mathrm{erg}\left(\frac{F_{\mathrm{jb}}}{\mathrm{nJy}}\right)^{0.63}\left(\frac{t_{\mathrm{jb}}}{\mathrm{day}}\right)^{1.10}\left(\frac{d_{L}}{\mathrm{Gpc}}\right)^{1.25} & \mathrm{infrared}\:\left(\zeta\times10^{14}\,\mathrm{Hz}\right)\\
\left(4.1\pm1.2\pm0.5\right)\times10^{49}\,\mathrm{erg}\left(\frac{F_{\mathrm{jb/max}}}{\mathrm{\mu Jy}}\right)^{0.18}\left(\frac{t_{\mathrm{jb/max}}}{\mathrm{day}}\right)^{1.59}\left(\frac{d_{L}}{\mathrm{Gpc}}\right)^{0.35} & \mathrm{radio}\:\left(\zeta\times10^{10}\,\mathrm{Hz}\right)
\end{cases}\label{eq:E-on-2-parameters}
\end{align}

\begin{align}
E_{K,\mathrm{iso}}^{\mathrm{\left(off\right)}} & =\begin{cases}
\left(1.6\pm0.4\pm0.3\right)\times10^{51}\,\mathrm{erg}\left(\frac{F_{\mathrm{max}}}{\mathrm{nJy}}\right)^{0.72}\left(\frac{t_{\mathrm{max}}}{\mathrm{day}}\right)^{1.17}\left(\frac{d_{L}}{\mathrm{Gpc}}\right)^{1.44} & \mathrm{x-ray}\:\left(\zeta\times10^{17}\,\mathrm{Hz}\right)\\
\left(6.0\pm2.3\pm0.9\right)\times10^{50}\,\mathrm{erg}\left(\frac{F_{\mathrm{max}}}{\mathrm{nJy}}\right)^{0.63}\left(\frac{t_{\mathrm{max}}}{\mathrm{day}}\right)^{1.03}\left(\frac{d_{L}}{\mathrm{Gpc}}\right)^{1.26} & \mathrm{infrared}\:\left(\zeta\times10^{14}\,\mathrm{Hz}\right)\\
\left(2.1\pm1.0\pm0.5\right)\times10^{49}\,\mathrm{erg}\left(\frac{F_{\mathrm{max}}}{\mathrm{\mu Jy}}\right)^{0.41}\left(\frac{t_{\mathrm{max}}}{\mathrm{day}}\right)^{1.66}\left(\frac{d_{L}}{\mathrm{Gpc}}\right)^{0.82} & \mathrm{radio}\:\left(\zeta\times10^{10}\,\mathrm{Hz}\right)
\end{cases}\label{eq:E-off-2-parameters}
\end{align}

\end{widetext}

As can be seen this removes the main uncertainty in the analysis and
leaves mostly the uncertainty from the linear regression. It is quite
striking that because of the simpler numerical procedure in this case,
merely the position of the jet-break respectively maximum gives effectively
even more precise estimates with merely a few tens of percent uncertainty,
that are in particular not affected by the fact that realistic light
curves have a more complicated form compared to the simple light curve
parameterizations used before. As can be seen, the same spectral flux
density in the off-axis case implies/requires a higher kinetic energy
compared to the on-axis case, which is consistent with expectations.
Similarly, the same spectral flux density at higher frequencies implies/requires
a higher kinetic energy, confirming that jet-break happens at a late
time in the process where most of the emission is already low energy
\footnote{Note the different characteristic spectral flux density of $\mathrm{\mu Jy}$
in the radio case compared to $\mathrm{nJy}$ in the other bands.}. Moreover, the power-law dependencies are remarkably similar in the
on- and the off-axis case at the higher frequencies. Taking into account
that these light curves describe the same events seen at different
angles, that the considered characteristic feature in the light curve
stems in both cases from the same underlying mechanism, and that both
eqs. (\ref{eq:E-on-2-parameters}) and (\ref{eq:E-off-2-parameters})
individually already show that the kinetic energy can be predicted
irrespective of the observing angle, this should be less surprising
and presents another consistency check of our results. The maximum
in the on-axis light curve at radio frequencies involves additional
physics due to the spectral evolution, and therefore it is likewise
consistent that the power law dependencies are more different in this
case.

Note that the expressions eqs. (\ref{eq:E-on-4-parameters}) to (\ref{eq:E-off-2-parameters})
are qualitatively different from similar expressions given in the
literature \citep{Zhang:2006uj,Fong:2015oha} that determine the \emph{kinetic
energy at any time in terms of the spectral flux density at that time}
and various input parameters of the forward shock model. Eqs. (\ref{eq:E-on-4-parameters})
to (\ref{eq:E-off-2-parameters}), in contrast, determine the \emph{initial
kinetic energy in terms of the values at jet-break}, i.e. a much later
point in the evolution. They only depend on light curve parameters
while they do \emph{not} require further information. Nonetheless
it is interesting to note that the power law indices describing the
dependence on the spectral flux density and on time obtained in \citep{Zhang:2006uj}
for the average value of $p=2.43$ considered are similar to those
we find here. 

Finally we consider whether the energy can even be predicted with
merely the spectral flux density $F_{\mathrm{max/jb}}$ alone. As
seen in the last row of tables \ref{tab:Weights-and-biases-energy-x-ray}
to \ref{tab:Weights-and-biases-energy-radio}, the relative errors
increase strongly in this case which clearly shows that a single parameter
is not enough to predict the kinetic energy. This is consistent with
previous work, e.g. \citep{2007MNRAS.377.1464N} showing that the
time of jet-break is correlated with but does not fix the kinetic
energy. Correspondingly, this means that the time and the flux are
not clearly correlated and both of them contain independent information
that is required. However, observationally this is surely no problem
at all since whenever the jet break in the on-axis case respectively
the maximum in the off-axis case can be located in the light curve,
both the time and the flux are known anyway.

\section{Conclusions}

We establish a simple quantitative connection between input parameters
of the forward shock model and the morphology of the corresponding
afterglow light curve. It allows to obtain approximate quantitative
insight into the energetics of the emission mechanism directly from
a few physical parameters that can be readily read off a given light
curve. Strikingly we find that even merely a single point on the light
curve---namely the position of the jet-break respectively maximum---is
sufficient to obtain a reliable estimate for the equivalent isotropic
kinetic energy, partly characterizing the strength of the GRB central
engine, at the tens of percent level. This uncertainty is at the same
level than many other uncertainties in the process and therefore our
expressions present a dramatic yet viable simplification. This presents
the main result of our work and shows that although the dynamical
mechanisms and the resulting afterglow light curves are rich and complex,
their key features are nonetheless determined by simple energetics.
Moreover, it shows that the different light curve parameters are strongly
correlated.

GRB light curves are complicated and include in addition to features
from the external shock, considered here, also signatures from various
other astrophysical mechanisms. This is in particular the case at
early times where both internal effects, such internal plateaus and
flares, attributed to a long-lasting central engine, or an additional
external reverse shock are important \citep{Zhang:2018ond}. Reverse
shocks can be strongly magnetized and are known to affect some observed
optical and radio light curves. Although these other light curve features
apparently stem from different emission mechanisms and sites, they
can overshadow the forward shock signal and are not included in our
analysis. Therefore, our results and in particular the estimate of
the strength of the central engine based on the entire light curve
should only be applied to sufficiently ``clean'' light curves that
show a ``canonical'' form and are e.g. not strongly affected by
flare activity, so that the relevant light curve features can be clearly
extracted. Similarly, some short GRBs show extended high-energy emission
\citep{Kaneko:2015lga}, which could likewise affect the early time
afterglow. The fact that even the position of the jet-break respectively
the maximum is enough to determine the kinetic energy circumvents
these problems. This is even more important taking into account that
in addition to the GRB there is often also emission from a kilonova
\citep{Metzger:2016pju}, that strongly dominates initially in the
infrared and optical band for off-axis sources \citep{Barnes:2020nfi},
so it is very favorable that no early time information is needed to
learn about the energetics of the process.

Our present study covers key parameters of the forward shock model
while other parameters have so far been set to fiducial values. These
are the jet opening angle $\mathrm{\theta_{\mathrm{jet}}}$, the fraction
of the jet energy in magnetic fields $\epsilon_{m}$ and electrons
$\epsilon_{e}$ respectively, and the electron spectral index $p$.
Since we compute with the isotropic kinetic energy effectively a local
quantity ($4\pi$ times the energy per solid angle) and the jet-break
depends on the gamma factor of electrons locally somewhere in the
jet becoming too small, the actual size of the jet should (aside from
boundary effects) not significantly affect this relation. Similarly
since most afterglows are well described by a weakly magnetized forward
shock, the precise value of the magnetization should likewise not
significantly affect the light curve as long as it is sufficiently
weak. The energies carried by the different components can be written
in terms of the energy fractions $E_{e/p,\mathrm{iso}}=\epsilon_{e/p}E_{K,\mathrm{iso}}$
and neglecting the magnetic component yields $\epsilon_{m}=0$, $\epsilon_{p}=1-\epsilon_{e}$.
The electrons strongly dominate the synchrotron luminosity $L_{\mathrm{syn}}$
due to their large Lorentz factor and assuming that the composition
stays constant, since the baryonic component in a plasma is strongly
electrically coupled and replenishes part of the lost energy, gives

\begin{equation}
\frac{dE_{e,\mathrm{iso}}}{dt}=\epsilon_{e}\frac{dE_{K,\mathrm{iso}}}{dt}=-\left(1-\epsilon_{e}\right)\frac{dE_{K,\mathrm{iso}}}{dt}-L_{\mathrm{syn}}
\end{equation}
which shows that the terms involving $\epsilon_{e}$ cancel, and the
evolution of $E_{K,\mathrm{iso}}$, and correspondingly also its starting
value, is in this approximation not explicitly dependent on the electron
fraction \footnote{The luminosity, respectively spectral flux density, in principle implicitly
depends on the electron fraction, but our expressions for the isotropic
kinetic energy in terms of light curve parameters eqs. (\ref{eq:E-on-4-parameters})-(\ref{eq:E-off-2-parameters})
explicitly depend on the spectral flux density.}. Finally, while we used here the average value for the spectral index
$p$ estimated from observed GRBs, this parameter could in principle
have a larger impact. Theoretical predictions suggest that it can
affect the power law exponents of the other quantities \citep{Zhang:2018ond}.
This cannot be captured by our simple linear model and including this
dependence will require a more sophisticated analysis, e.g. via a
deep neutral network \footnote{Unfortunately, such an analysis would not provide such simple analytic
relations we obtain here.}.

Similarly we only consider the simplest jet shape given by a ``top-hat''
jet. GRB170817 highlighted the importance of structured jets \citep{Ryan:2019fhz,Ilhan:2019bmg}
for certain nearby GRB events that can even be observed off-axis.
While a ``top-hat'' jet, being the simplest choice, works well for
most observed distant GRBs that can generally only be observed nearly
on-axis, off-axis events are relevant for multi-messenger observations
and our parametrization could be extended in the future to include
more complicated jet shapes. As already discussed, while the applied
forward shock model assumes a weakly magnetized forward shock, some
GRB afterglows seem to require larger magnetization, so that our results
do not directly apply in these cases \citep{Zhang:2018ond}.

The motivation and long term goal of our work is to use the afterglow
light curve morphology to learn about the neutron star merger dynamics
from these at present exclusively available late-time electromagnetic
signals. Even though merger events are immensely complex, tracking
conserved quantities, like energy and angular momentum, could eventually
allow a connection of these distant stages. The isotropic kinetic
energy studied here yields in combination with the jet opening angle
$\theta_{\mathrm{jet}}$, which can be obtained independently \citep{Frail:2001qp},
the total kinetic energy of the afterglow $E_{K}=\left(1-\cos\theta_{\mathrm{jet}}\right)E_{K,\mathrm{iso}}$.
The central engine resulting from the merger product powers both the
observed prompt GRB emission and the afterglow, and adding these energies
$E_{\mathrm{CE}}=E_{\gamma}+E_{K}$ provides an estimate for the total
energy output of the central engine. In contrast to the afterglow,
the prompt emission is very irregular and variable \citep{Zhang:2018ond}
and the GRB efficiency $\eta_{\gamma}\equiv E_{\gamma,\mathrm{iso}}/E_{\mathrm{CE,iso}}$
can observationally vary drastically ranging from less than 1\% to
more than 90\% \citep{Zhang:2006uj,Fong:2015oha}. This points to
a turbulent process with stochastically varying emission, which is
expected in the scenario of an internal shock, where the jet smashes
into the inherently strongly inhomogeneous and anisotropic ejecta
resulting from the merger. In case the emission is only temporally
varying the observationally obtained value $E_{\mathrm{CE}}$ presents
a good estimate for the total energy output of the central engine.
Yet, in case the emission is also spatially stochastically varying
(e.g. since the emission is locally obscured by other debris), what
fraction of the prompt emission we observe could be to some extent
random. In this case the observed gamma ray flux might not be a reliable
measure for the strength of the central engine, and the actual efficiencies
of GRBs might vary much less than what is observed. The much more
continuous and gradually varying late time afterglow in contrast presents
a much more controlled data set, that, as our work shows, indeed directly
encodes information about the external energy injected by the central
engine. Thereby, the present work covers a first link in a connection
to the earlier stages of the merger process \citep{Baiotti:2016qnr}.
\begin{acknowledgments}
We are grateful to Massimiliano de Pasquale for very helpful discussions.
This work was supported by the Turkish Research Council (TÜBITAK)
via project 119F073.
\end{acknowledgments}

\bibliographystyle{apsrev}
\bibliography{references}

\end{document}